\documentstyle[aps,twocolumn,eqsecnum]{revtex}
\begin{document}
\input epsf

\wideabs{
\title{Can the post-Newtonian gravitational waveform of an inspiraling
binary be improved by solving the energy balance equation
numerically?} 

\author{Wolfgang Tichy and \'Eanna \'E. Flanagan}
\address{Center for Radiophysics and Space Research,
Cornell University, Ithaca, NY 14853}
\author{Eric Poisson}
\address{Department of Physics, University of Guelph, Guelph, Ontario,
N1G 2W1, Canada}  

\date{\today}

\maketitle

\begin{abstract}
The detection of gravitational waves from inspiraling compact binaries 
using matched filtering depends crucially on the availability of
accurate template waveforms.  We determine whether the accuracy of the
templates' phasing can be improved by solving the post-Newtonian
energy balance equation numerically, rather than (as is normally done)
analytically within the post-Newtonian perturbative expansion.  By
specializing to the limit of a small mass ratio, we find evidence that
there is no gain in accuracy.
\end{abstract}
}
\def\beq{\begin{equation}}
\def\endeq{\end{equation}}

\narrowtext

\section{Introduction and Summary}

Several kilometer-scale interferometric gravitational wave
detectors are currently being built, among them the two American LIGO
detectors, the French-Italian VIRGO detector, the German-British GEO
600 detector and the Japanese TAMA detector. Gravitational waves from
inspiraling compact binaries are among the most promising candidates
to be detected. In order to extract a gravitational wave signal from
the noisy background the technique of matched filtering
\cite{Wainstein,Finn} will be used. One of the drawbacks of matched
filtering is that the theoretical templates used must be close to the
actual gravitational wave signal in order to detect the signal and
estimate its parameters. In the case of a nearly circular inspiral of
two point masses the expected gravitational wave signal has the form
of a chirp, i.e., a roughly sinusoidal signal with gradually
increasing amplitude and frequency. If such a signal is to be detected
by matched filtering, high accuracy is needed in the templates; in
particular, the phase of the template must closely match the phase of
the actual signal. Inspiral templates have been calculated to date up
to post-2.5-Newtonian order \cite{Blanchet}.   

Several authors have investigated the question of to what
post-Newtonian order does one need to push template computations in
order to have an acceptably small template-inaccuracy reduction in
event detection rate \cite{accuracy_required,paperVI,Droz,haris}.  The
result is that post-2-Newtonian templates may be sufficiently accurate
to detect neutron star/neutron star binaries \cite{note}; the loss in
event rate in this case is $\sim 12\%$ for initial LIGO and $\sim 20
\%$ for advanced LIGO \cite{Droz}. 

However, there are several motivations for trying to obtain more
accurate templates.  First, there may well be a high event rate of
neutron star/black hole or black hole/black hole inspirals.  For these
more massive systems the accuracy requirements are more stringent,
since the frequency band ($\sim 50 - 200 \, {\rm Hz}$) where most of the
signal-to-noise is accumulated is in a more relativistic regime for
more massive systems.  For example, for initial LIGO detectors and for
a binary system of a $4 M_\odot$ black hole and a $30 M_\odot$ black
hole, using post-2-Newtonian search templates would allow us to detect
only $\sim 35 \%$ of the signals that otherwise would be detectable
\cite{Droz}.  A second motivation is that one will need high accuracy
templates in order to avoid appreciable systematic errors in measurements
of the binary's parameters \cite{DIS,Sathya,BSD}.    

A variety of methods of increasing template accuracy have been pursued
in recent years.  First, one can compute the
templates up to ever-higher post-Newtonian orders; this is arduous but
going beyond the current post-2.5-Newtonian templates is feasible.
Progress is being made on computing post-3-Newtonian templates; see
Ref.\ \cite{Kip} and references therein. 
Second, a celebrated result in this field was the discovery by Damour,
Iyer, and Sathyaprakash \cite{DIS} that using Pad\'e approximants can
significantly increase the accuracy of template phasing.  Third,
Damour and Buonanno \cite{DB} have suggested a particular ansatz for 
obtaining templates containing additional terms of all post-Newtonian
orders, starting, say, from post-2-Newtonian templates.

The purpose of this paper is to investigate yet another possible
method of improving the accuracy of post-Newtonian templates.  
The basic idea is very simple.  In computing post-2-Newtonian
templates, for example, one should strictly speaking 
discard all terms of post-2.5-Newtonian order (and higher)
everywhere in the calculations.  To do otherwise would be
inconsistent with the perturbation expansion method.  Yet, there could 
be pieces of the calculation for which {\it retaining}
post-2.5-Newtonian (and higher) order terms would lead to improved
accuracy.  For example, the dominant, $m=2$
piece of the waveform is usually written as
\begin{equation}
h(t) = A(t) \cos \left[ \phi^{GW}(t) \right],
\label{basic1}
\end{equation}
where both the amplitude $A(t)$ and the phase $\phi^{GW}(t)$ have
separate post-Newtonian expansions, $A = \sum_j \varepsilon^j A^{(j)}$
and $\phi^{GW} = \sum_j \varepsilon^j \phi^{(j)}$, with $\varepsilon$
a formal expansion parameter.  Now a perturbation theory purist would
insist on inserting the expansion for $\phi^{GW}$ into Eq.\
(\ref{basic1}) and on expanding the cosine using a Taylor expansion.
However, it is well known that the resulting expression is a much
poorer representation of the true signal than the original un-expanded
form (\ref{basic1}).

The question then arises: are there other stages in the construction
of post-Newtonian templates where one discards higher order terms,
which, if retained, might lead to increased accuracy?  A natural
possibility is the stage in which one goes from the post-Newtonian
formulae for the energy flux $F(f) = -dE / dt(f)$ and orbital energy $E(f)$
as functions of
gravitational wave frequency $f$, to the formula for the phase
$\phi^{GW}(t)$ of the gravitational waveform.  Given analytical
formulae for $F(f)$ and $E(f)$ up to some post-Newtonian order, one
can either (I) 
solve analytically for $\phi^{GW}(t)$ within the post-Newtonian
approximation, discarding all higher order terms, or (II) one can {\it
numerically solve} the energy balance equation to obtain
$\phi^{GW}(t)$.  This second procedure effectively generates and
retains terms at all post-Newtonian orders, so is strictly speaking
inconsistent, but one might hope that it would lead to increased
accuracy.  We note that the papers \cite{accuracy_required,paperVI,DIS}
investigating the accuracy of post-Newtonian templates have generally
used method (II), whereas the popular data analysis package GRASP
\cite{GRASP} used in Ref.\ \cite{upperlimit} uses method (I). 
The GRASP manual \cite{GRASP} speculates that method (II) might be
more accurate.  

In this paper we present evidence, based on the limiting case of
binaries with small mass ratios, that numerically solving the energy 
balance equation does not in fact increase the accuracy.  
We arrive at this conclusion after checking the accuracies of methods
(I) and (II) in three different ways. We compare expansion
coefficients of the Fourier transform of the waveform; we numerically
find the relative error in the Fourier transform of the waveform; and
we compute overlaps of templates constructed from both methods with
the exact waveform.  
The result, that the numerical solution of the energy balance
equation (method (II)) does not increase the accuracy,
is disappointing, but constitutes useful information from the point of
view of generating template banks for inspiral searches: there is no
motivation in terms of increased event rate to solve numerically for
the wave's phasing.

\section{Method of calculation}

In order to explain our calculation, we first summarize how the
waveform's phasing can be computed from the energy flux function
$F(f)$ and orbital energy function $E(f)$, where $f$ is gravitational
wave frequency \cite{DIS,paperVI}.
Let $m_1$, $m_2$ be the masses of the two components
of the binary and  $m = m_1 + m_2$ be the total mass.
Let $\phi(t)$ be
the orbital phase of the binary, so that $\phi^{GW}(t) = 2 \phi(t)$,
where $\phi^{GW}$ is the phase of the dominant, $m=2$ piece of the
waveform.  We define the dimensionless variable
\beq
v = (\pi m f)^{1/3}.
\label{V-def}
\endeq
[Here and throughout we use units with $G=c=1$.]
The orbital phase $\phi(t)$ is derived from the relation
\begin{equation}                                     
\label{phi-omega} 
 \frac{d \phi}{dt} = \pi f,
\end{equation}
and from the energy balance equation
\begin{equation}
\label{E-Balance}
\frac{d E(v)}{dt} = -F(v).
\end{equation}
Equations (\ref{V-def}) -- (\ref{E-Balance}) 
yield a parametric solution for $\phi(t)$ given by
\begin{equation}
\label{phi_V}   
\phi(v) = \phi_c - {1 \over m} \int^v_{v_i} d {\bar v} {\bar v}^3 \,
{E'({\bar v}) \over F({\bar v})}, 
\end{equation}
and
\begin{equation}
\label{t_V}   
t(v) = t_c - \int^v_{v_i} d {\bar v} { E'({\bar v}) \over F({\bar v})},
\end{equation}
where $\phi_c$, $t_c$ and $v_i$ are constants.  

In the restricted post-Newtonian approximation, in which we neglect
the $m\ne2$ multipoles, the gravitational waveform has the form
$h(t) = A(t)\cos[\phi^{GW}(t)]$, where $A(t)$ is a slowly varying
amplitude.  The Fourier transform ${\tilde h}(f)$ of this waveform is 
$
\tilde{h}(f) = B(f) \mbox{e}^{i \psi(f) },
$
where $ B(f)$ is some frequency dependent prefactor, and where, in the
stationary phase approximation, the phase $\psi(f)$ is given by   
\begin{equation}
\label{psi_stat_phase}
\psi(f)=2\pi f t(v) -2\phi(v) -\frac{\pi}{4}.
\end{equation}
Using Eqs.\ (\ref{phi_V}) and (\ref{t_V}) gives the frequency
domain phase \cite{paperVI}:
\begin{eqnarray}
\psi(f) &=& 2 (t_c/m) v^3 - 2 \phi_c - \pi/4 \nonumber \\
\mbox{} && - {2 \over m} \int_{v_i}^v d {\bar v} \ (v^3 - {\bar v}^3)
\, { E^\prime({\bar v}) \over F({\bar v}) }.
\label{st}
\end{eqnarray}
We note that the corrections to the stationary phase approximation are
very small, arising only at post-5-Newtonian order
\cite{Droz-K-Poisson-O}, so it is sufficient for our purposes to use
the expression (\ref{st}). 

Equation (\ref{st}) is the starting point for our analysis.  
We will investigate the accuracy with which various approximations 
reproduce the Fourier-domain phase $\psi(f)$, which is the version of  
the phase function that is most relevant for matched filtering. 
The two possible calculational methods we consider are (I) 
to insert post-Newtonian expressions for the functions $E(v)$ and
$F(v)$ into Eq.\ (\ref{st}), and discard all the higher order
post-Newtonian terms generated, and (II) to insert post-Newtonian
expressions for the functions $E(v)$ and $F(v)$ into Eq.\ (\ref{st})
and solve exactly for the phase $\psi(f)$, retaining all the
higher order post-Newtonian terms generated.  

To assess the accuracy of each of these two methods, we specialize to
the limit $m_1 m_2/m^2 \to 0$ for which the functions $E(v)$ and
$F(v)$ are known \cite{paperVI,DIS,Tanaka-Tagoshi-Sasaki}.   
We then check the accuracy of method (I) and (II) in three ways.

\subsection{Checking the accuracy of methods (I) and (II) 
by comparing expansion coefficients of $\psi(f)$}

The first check is entirely analytical. We expand all the phase
functions $\psi(f)$ as post-Newtonian power series in $v$ up to some
high order (e.g. post-5.5-Newtonian), and 
compare the accuracy with which methods (I) and (II) reproduce the
coefficients in this power series.   While this comparison procedure
is less accurate than comparing the phases produced by methods (I) and
(II) to the exact numerical phase, it does allow us to check whether 
there is any indication that method (II) is more accurate than method
(I).

In more detail, our comparison procedure works as follows. The
post-Newtonian expansions for the functions $E(v)$ and $F(v)$ have the
general form 
\begin{equation}
\label{E_V}
E(v) =-{1 \over 2} \eta m  v^2 \left[1 + \sum_{i=1}  e_{2i} v^{2i} \right] ,
\end{equation}
\begin{equation}
\label{F_V}
F(v) =\frac{32 }{5}\eta^2 v^{10} 
\left[1 + \sum_{i=2} f_i v^i    
+ \sum_{i=6} g_i \mbox{ln} (v) v^i +\ldots\right] ,
\end{equation}
where $\eta = m_1 m_2/ m^2$ is the dimensionless mass ratio.
The ellipses in Eq.\ (\ref{F_V}) represent possible terms proportional
to $(\ln v)^m$ for $m \ge 2$ which could arise at high
post-Newtonian orders.  
The coefficients $e_i$, $f_i$ and $g_i$ in Eqs.\ (\ref{E_V}) and
(\ref{F_V}) are functions of the mass ratio $\eta$.  For general mass
ratios, the coefficients $e_i$ and $f_i$ in
are known up to $e_4$ and $f_5$ \cite{Blanchet}, while for $\eta=0$
all the $e_i$ coefficients are known \cite{paperVI,DIS} and the $f_i$ and
$g_i$ coefficients are known up to $f_{11}$ and $g_{11}$
\cite{Tanaka-Tagoshi-Sasaki}.  The known 
coefficients are tabulated in Appendix \ref{sec-e_f_B}.

If we now insert the expansions (\ref{E_V}) and (\ref{F_V}) into the
formula (\ref{st}) for the phase $\psi(f)$ we obtain
\begin{equation}
\label{psi_P}
\psi(f)=\frac{3v^{-5}}{128\eta} 
         \Bigg[ P(v)+\frac{256\eta}{3 m} v^8 t_{K} 
                    +\frac{128\eta}{3} v^5 K \Bigg] . 
\end{equation}
Here $t_{K}$ and $K$ are constants which correspond to the initial time 
and initial phase, and the function $P(v)$ has the expansion
\begin{equation}
\label{P_V}
P(v) = 1 + \sum_{j=2} \left\{ p_j  
+q_{j} \mbox{ln}(v)  +r_{j} [\mbox{ln}(v)]^2 + \ldots \right\} v^j .
\label{PofV}
\end{equation}
The coefficients $p_j$, $q_j$ and $r_j$ in Eq.\ (\ref{P_V}) 
are functions of the coefficients $e_{i}$, $f_{i}$ and $g_{i}$ in 
Eqs.\ (\ref{E_V}) and (\ref{F_V}) for $i \le j$:
\begin{equation}
p_j = p_j (e_1, \ldots e_j , f_1, \ldots f_j, g_1, \ldots g_j),
\end{equation}
\begin{equation}
q_j = q_j (e_1, \ldots e_j , f_1, \ldots f_j, g_1, \ldots g_j),
\end{equation}
\begin{equation}
r_j = r_j (e_1, \ldots e_j , f_1, \ldots f_j, g_1, \ldots g_j).
\end{equation}
For example, the expressions for the first few $p_j$'s are 
\begin{equation}
p_{2}= \frac{20(2e_2 -f_2)}{9} ,
\label{peg2}
\end{equation}
\begin{equation}
p_{3} = -4f_3 ,
\end{equation}
and
\begin{equation}
p_{4} =10(f_{2}^2 -2e_2 f_2 +3e_4 -f_4) .
\label{peg4}
\end{equation}

Suppose now that the functions $E(f)$ and $F(f)$ are known to
post-$N$-Newtonian order.  Then the coefficients $e_i$, $f_i$ and
$g_i$ are known for $0 \le i \le 2 N$.  If we now follow the usual
method (I) to generate the phase function $\psi(f)$, we obtain an
expansion of the form (\ref{P_V}) where the coefficients are given 
by
\begin{equation}
\label{p-methodI}
{}^{(I)}_{\ N}p_j = \left\{ \begin{array}{ll} 
p_j (e_1, \ldots e_j , f_1, \ldots f_j, g_1, \ldots g_j)
 & \mbox{ $j \le 2N$,} \nonumber\\ 
0 & \mbox{
        $j > 2N$,}\\ \end{array} \right.
\end{equation}
\begin{equation}
\label{q-methodI}
{}^{(I)}_{\ N}q_j = \left\{ \begin{array}{ll} 
q_j (e_1, \ldots e_j , f_1, \ldots f_j, g_1, \ldots g_j)
 & \mbox{ $j \le 2N$,} \nonumber\\ 
0 & \mbox{
        $j > 2N$,}\\ \end{array} \right.
\end{equation}
and
\begin{equation}
\label{r-methodI}
{}^{(I)}_{\ N}r_j = \left\{ \begin{array}{ll} 
r_j (e_1, \ldots e_j , f_1, \ldots f_j, g_1, \ldots g_j)
 & \mbox{ $j \le 2N$,} \nonumber\\ 
0 & \mbox{
        $j > 2N$.}\\ \end{array} \right.
\end{equation}
Here the superscript (I) means method (I) and the subscript $N$ refers
to the post-$N$-Newtonian approximation.  On the other hand, if we use
instead the method (II) to generate 
$\psi(f)$, we obtain an expansion with expansion coefficients
\begin{equation}
\label{p-methodII}
{}^{(II)}_{\ \ N}p_j = \left\{ \begin{array}{ll} 
p_j (e_1, \ldots e_j , f_1, \ldots f_j, g_1, \ldots g_j)
 & \mbox{ $j \le 2N$,} \nonumber\\ 
p_j (e_1, \ldots e_{2 N} , 0, \ldots 0, f_1, \ldots f_{2 N}, 
& \nonumber \\
\mbox{\ \ \ \ \ \ \ \ \ }0,\ldots 0, g_1, \ldots g_{2N}, 0, \ldots 0)
& \mbox{  $j > 2N$,}\\ \end{array} \right.
\end{equation}
\begin{equation}
\label{q-methodII}
{}^{(II)}_{\ \ N}q_j = \left\{ \begin{array}{ll} 
q_j (e_1, \ldots e_j , f_1, \ldots f_j, g_1, \ldots g_j)
 & \mbox{ $j \le 2N$,} \nonumber\\ 
q_j (e_1, \ldots e_{2 N} , 0, \ldots 0, f_1, \ldots f_{2 N}, 
& \nonumber \\
\mbox{\ \ \ \ \ \ \ \ \ }0,\ldots 0, g_1, \ldots g_{2N}, 0, \ldots 0)
& \mbox{  $j > 2N$,}\\ \end{array} \right.
\end{equation}
and
\begin{equation}
\label{r-methodII}
{}^{(II)}_{\ \ N}r_j = \left\{ \begin{array}{ll} 
r_j (e_1, \ldots e_j , f_1, \ldots f_j, g_1, \ldots g_j)
 & \mbox{ $j \le 2N$,} \nonumber\\ 
r_j (e_1, \ldots e_{2 N} , 0, \ldots 0, f_1, \ldots f_{2 N}, 
& \nonumber \\
\mbox{\ \ \ \ \ \ \ \ \ }0,\ldots 0, g_1, \ldots g_{2N}, 0, \ldots 0)
& \mbox{  $j > 2N$.}\\ \end{array} \right.
\end{equation}
Thus, the two methods agree on $p_j$, $q_j$ and $r_j$ for $j \le 2 N$,
but for $j > 2N$ method (I) gives expansion coefficients of zero while
method (II) yields coefficients of the form $p_j (e_1, \ldots e_{2 N}
, 0, \ldots 0, f_1, \ldots f_{2 N}, 0,\ldots 0, g_1, \ldots g_{2N}, 0,
\ldots 0)$ which differ somewhat from the true values $p_j (e_1,
\ldots e_{j}, f_1, \ldots f_{j}, g_1, \ldots g_{j})$
because of having the coefficients $e_i$, $f_i$ and $g_i$ set to zero for
$2N+1 \le i \le j$.  We define 
\begin{equation}
p_{j,k} = {}^{(II)}_{\ k/2}p_j ,
\end{equation} 
and similarly for $q_j$ and $r_j$, so that
$p_{k,k} = p_k$.

As an example, suppose that the functions $E(f)$ and $F(f)$ were known
only up to post-1.5-Newtonian order, so that only the coefficients
$e_2$, $f_2$ and $f_3$ were known, but not $e_4$ and $f_4$.  
Up second post-Newtonian order the expansion
(\ref{P_V}) has the form
\begin{equation}
P(v)=1+p_2 v^2 +p_3 v^3 +p_4 v^4 +O(v^5), 
\end{equation}
where the coefficients $p_2$, $p_3$ and $p_4$ are given in 
Eqs.\ (\ref{peg2}) -- (\ref{peg4}) above.  
How accurately could
we determine the coefficients $p_2$, $p_3$ and $p_4$ in this case?
Obviously we could compute $p_{2}$ and $p_{3}$ exactly since they do 
not depend on $e_4$ and $f_4$.  However, the coefficient $p_{4}$ does
depend on $e_4$ and $f_4$, and can be written as [cf.\ Eq.\
(\ref{peg4}) above]
\begin{equation}
p_{4}= p_{4,3} + \Delta p_{4,3}.
\label{psplit}
\end{equation}
Here 
\begin{equation}
p_{4,3} = {}^{(II)}_{\ 1.5}p_4= 10 \, (f_{2}^2 -2e_2 f_2) ,
\label{p43}
\end{equation}
is the piece of $p_4$ that can be computed from the post-1.5-Newtonian
pieces of $E(f)$ and $F(f)$; it is thus nonlinear in the coefficients
$e_2$ and $f_2$.  The value (\ref{p43}) is the prediction of method
(II), while the 
method (I) gives instead the value ${}^{(I)}_{1.5}p_4 =0$.
The error term in Eq.\ (\ref{psplit}) is given by
\beq
\Delta p_{4,3} =10 \, (3e_4 -f_4)
\endeq
and is linear in the post-2-Newtonian coefficients $e_4$ and
$f_4$.  Using the values of $e_2$, $f_2$, $e_4$ and $f_4$ listed in 
Appendix \ref{sec-e_f_B} we find that $\Delta p_{4,3} / p_4 \approx
-1.73$ for $\eta = 0$, which is rather large.
Hence in this particular example we do not improve the accuracy in
the coefficient $p_4$ by using method (II) rather than method (I).

In general, the question we want to address is whether 
the approximate coefficient ${}^{(II)}_{\ \ N}p_j = p_{j,2N}$ is typically
significantly closer to the true coefficient $p_j$ than zero is to
$p_j$, for $j > 2 N$, i.e., whether 
\beq
\frac{| p_{j,2N} - p_j |}{p_j} \alt ({\mbox {\rm a few tens of percent} })
\label{isok}
\endeq
(and similarly for $q_j$ and $r_j$).  
In Tables \ref{p_jk-table}, \ref{q_jk-table} and \ref{r_jk-table}
below we list the values of the 
true 
coefficients $p_j$, $q_j$ and $r_j$ and also the
approximate coefficients $p_{j,k}$, $q_{j,k}$ and $r_{j,k}$ for
various values of $k$, computed
from the values given in Appendix \ref{sec-e_f_B} using Eqs.\
(\ref{E_V}), (\ref{F_V}) and (\ref{st}).  We list the analytic
expressions for these approximate coefficients in Appendix 
\ref{sec-pqr}.  Examination of Tables \ref{p_jk-table}, \ref{q_jk-table} 
and \ref{r_jk-table} shows that there is no
tendency for the inequality (\ref{isok}) to be satisfied.  

Therefore method (II) does not seem to lead to a gain in 
accuracy when compared to method (I) in the test mass case 
($\eta \to 0$).

\onecolumn

\begin{table}
\caption{The coefficients 
$p_{j,k}={}^{(II)}_{k/2}p_j$ as calculated according to method (II).
These coefficients are what one obtains if the orbital energy $E(f)$
and gravitational wave luminosity $F(f)$ as functions of frequency $f$
are known only up to post-$k/2$-Newtonian order.  Note that the values
of $p_{j,k}$ differ significantly from their true values $p_j =
p_{j,j}$ for $k<j$.
\label{p_jk-table}}
\begin{tabular}{|l|rrrrrrrrrr|l|r|}
method (II) &k=2 & k=3 & k=4 & k=5 & k=6 & k=7 & k=8 & k=9 & k=10 & k=11 &
          \multicolumn{2}{l}{true values} \\
\hline
$p_{3,k}\times 10^{-3}$ 
&     0&  -0.0503& &&&& &&&& 
$p_{3}\times 10^{-3}$& -0.0503\\

$p_{4,k}\times 10^{-3}$ 
& 0.0821&  0.0821&  0.0301& &&&& &&&
$p_{4}\times 10^{-3}$& 0.0301 \\

$p_{5,k}\times 10^{-3}$ 
&    0&   0.331&   0.331&   0.161& &&&& &&
$p_{5}\times 10^{-3}$& 0.161 \\

$p_{6,k}\times 10^{-3}$ 
&  -0.609&  -3.77&  -3.60&  -3.60&   -0.441& &&&& &
$p_{6}\times 10^{-3}$& -0.441 \\

$p_{7,k}\times 10^{-3}$ 
&    0&  7.59&  7.52&  2.98&  2.98&   0.954& &&&& 
$p_{7}\times 10^{-3}$& 0.954\\

$p_{8,k}\times 10^{-3}$ 
&  -0.502&  -7.26&  -7.19&  -2.91&   0.828&   0.828&   0.995& &&& 
$p_{8}\times 10^{-3}$& 0.995\\

$p_{9,k}\times 10^{-3}$ 
&    0& -37.8& -40.2& -28.8&  5.61& 11.6& 11.6&  4.43& &&
$p_{9}\times 10^{-3}$& 4.43\\

$p_{10,k}\times 10^{-3}$
& 1.68& 43.3& 46.5& 15.2&  -1.76& -12.0& -11.5& -11.5&  -8.77& &
$p_{10}\times 10^{-3}$& -8.77\\

$p_{11,k}\times 10^{-3}$
&    0& -76.8& -82.5& -28.4& 19.4& 26.1& 23.9& 14.4& 14.4& 12.3& 
$p_{11}\times 10^{-3}$& 12.3\\ 
\end{tabular}
\end{table}

\begin{table}
\caption{The coefficients $q_{j,k}$; see caption
of Table \ref{p_jk-table}.
\label{q_jk-table}} 
\begin{tabular}{|l|rrrrrrrrrr|l|r|}
method (II) &k=2 & k=3 & k=4 & k=5 & k=6 & k=7 & k=8 & k=9 & k=10 & k=11 &
          \multicolumn{2}{l}{true values} \\
\hline
$q_{3 ,k}\times 10^{-3}$&    0&     0& &&&& &&&&
$q_{3}\times 10^{-3}$& 0\\

$q_{4 ,k}\times 10^{-3}$&    0&     0&     0& &&&& &&&
$q_{4}\times 10^{-3}$& 0\\

$q_{5 ,k}\times 10^{-3}$&    0& 0.992& 0.992& 0.482& &&&& &&
$q_{5}\times 10^{-3}$& 0.482\\

$q_{6 ,k}\times 10^{-3}$&    0&     0&     0&     0&-0.326& &&&& &
$q_{6}\times 10^{-3}$& -0.326\\

$q_{7 ,k}\times 10^{-3}$&    0&     0&     0&     0&     0&     0& &&&&
$q_{7}\times 10^{-3}$& 0\\

$q_{8 ,k}\times 10^{-3}$& 1.51&  21.8&  21.6&  8.74& -2.91& -2.91& -3.18&&&&
$q_{8}\times 10^{-3}$& -3.18\\

$q_{9 ,k}\times 10^{-3}$&    0&     0&     0&     0& -4.10& -4.10& -4.10& -2.05& &&
$q_{9}\times 10^{-3}$& -2.05\\

$q_{10,k}\times 10^{-3}$
&    0&     0&     0&     0&  1.95&  1.95&   0.702&   0.702&   0.235&&
$q_{10}\times 10^{-3}$& 0.235\\

$q_{11,k}\times 10^{-3}$
&    0&     0&     0&     0& -6.00& -6.00&  -3.05&   -0.356&   -0.356& -1.41&
$q_{11}\times 10^{-3}$& -1.41\\ 
\end{tabular}
\end{table}

\begin{table}
\caption{The coefficients $r_{j,k}$; see caption
of Table \ref{p_jk-table}.  These coefficients vanish for $j \le 7$.
\label{r_jk-table}}
\begin{tabular}{|l|rrrrrrrrrr|l|r|}
method (II) &k=2 & k=3 & k=4 & k=5 & k=6 & k=7 & k=8 & k=9 & k=10 & k=11 
&  
          \multicolumn{2}{l}{true values} \\
\hline
$r_{8,k}\times 10^{-3}$ & 0& 0& 0& 0& 1.29& 1.29& 0.584& &&&
$r_{8}\times 10^{-3}$ & 0.584\\

$r_{9,k}\times 10^{-3}$ & 0& 0& 0& 0&    0&    0&     0&  0& &&
$r_{9}\times 10^{-3}$ & 0\\

$r_{10,k}\times 10^{-3}$& 0& 0& 0& 0&    0&    0&     0&  0& 0& &
$r_{10}\times 10^{-3}$& 0\\

$r_{11,k}\times 10^{-3}$& 0& 0& 0& 0&    0&    0&     0&  0& 0& 0 &
$r_{11}\times 10^{-3}$& 0\\
\end{tabular}
\end{table}

\twocolumn

\subsection{Checking the accuracy of methods (I) and (II) numerically}

Next we perform a direct numerical check by comparing the phases
produced by methods (I) and (II) to the exact numerical phase. 
Note that the phase $\psi(f)$ in Eq. (\ref{st}) is not directly
observable, since it contains the unknown integration constants
$\phi_c$ and $t_c$, i.e. $\psi(f)$ is determined only up to a linear
function in $f$. Hence the relevant quantity to consider is
\begin{equation}
\psi'' (f)= \frac{d^2}{df^2} \psi (f)
=-\frac{2\pi^2 m}{3v^2}\frac{E'(v)}{F(v)} .
\end{equation}
In the test mass case $E'(v)$ and $F(v)$ are known 
exactly \cite{paperVI,DIS,Tanaka-Tagoshi-Sasaki},
and we can therefore find the exact $\psi''(f)$.

Now suppose $E'(v)$ and $F(v)$ are only known up to
post-$k/2$-Newtonian order. If method (I) is used $\psi'' (f)$ is
found to be 
\begin{equation}
{}^{(I)}_{k/2}\psi'' (f)
= -\frac{2\pi^2 m}{3v^2} \left[ \frac{E'(v)}{F(v)} \right]_k .
\end{equation}
On the other hand method (II) yields
\begin{equation}
{}^{(II)}_{k/2}\psi'' (f)
= -\frac{2\pi^2 m}{3v^2} \frac{\left[E'(v)\right]_k}{\left[F(v)\right]_k} . 
\end{equation}
Here $[...]_k$ denotes the powerseries of the expression inside the 
brackets with terms kept up to order $v^k$.

\begin{figure}
\epsfxsize=8.5cm 
\epsfbox{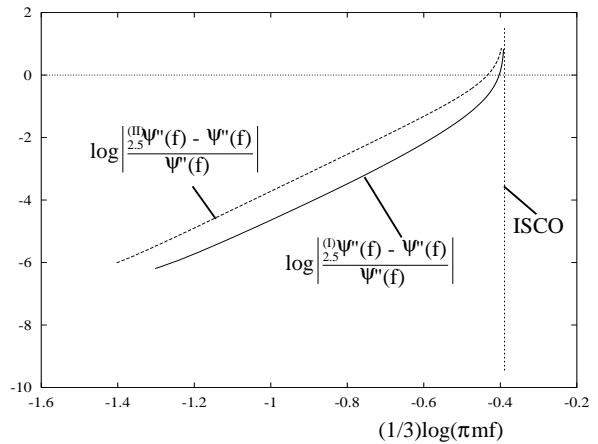}
\vspace{.4cm}
\caption{ 
The errors in the phase $\psi(f)$ of the Fourier transformed waveform  
produced by methods (I) and (II) in the test mass limit.
Plotted here are the logarithms of the relative errors in 
the second derivative $\psi''(f)$, for the case when the energy $E(f)$
and gravitational wave luminosity $F(f)$ are known only
up to post-2.5-Newtonian order.  The horizontal axis is ${\rm log}(\pi m f)/3$
where $m$ is the total mass of the system and $f$ is gravitational wave
frequency.  The line denoted ISCO indicates the 
location of the innermost stable circular orbit.
It can be seen that method (I) is more accurate for all frequencies $f$.  
}
\label{fig-PN2.5}
\end{figure}

\begin{figure}
\epsfxsize=8.5cm
\epsfbox{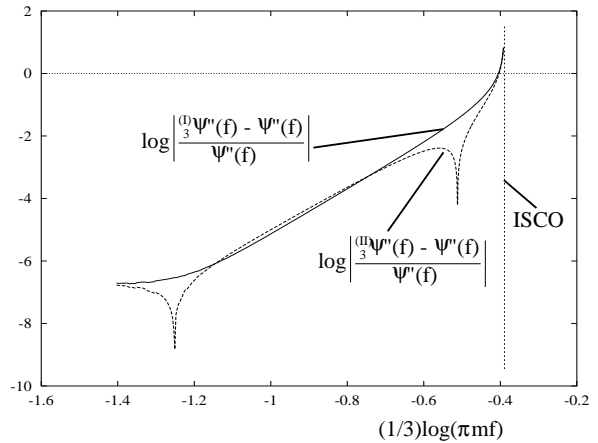}
\vspace{.4cm}
\caption{
The errors in the phase $\psi(f)$ when $E(f)$ and $F(f)$ are known
only up to post-3-Newtonian order; see caption of Fig.\ 1.
In this case method (II) is overall more accurate.
}    
\label{fig-PN3}
\end{figure}

\begin{figure}
\epsfxsize=8.5cm
\epsfbox{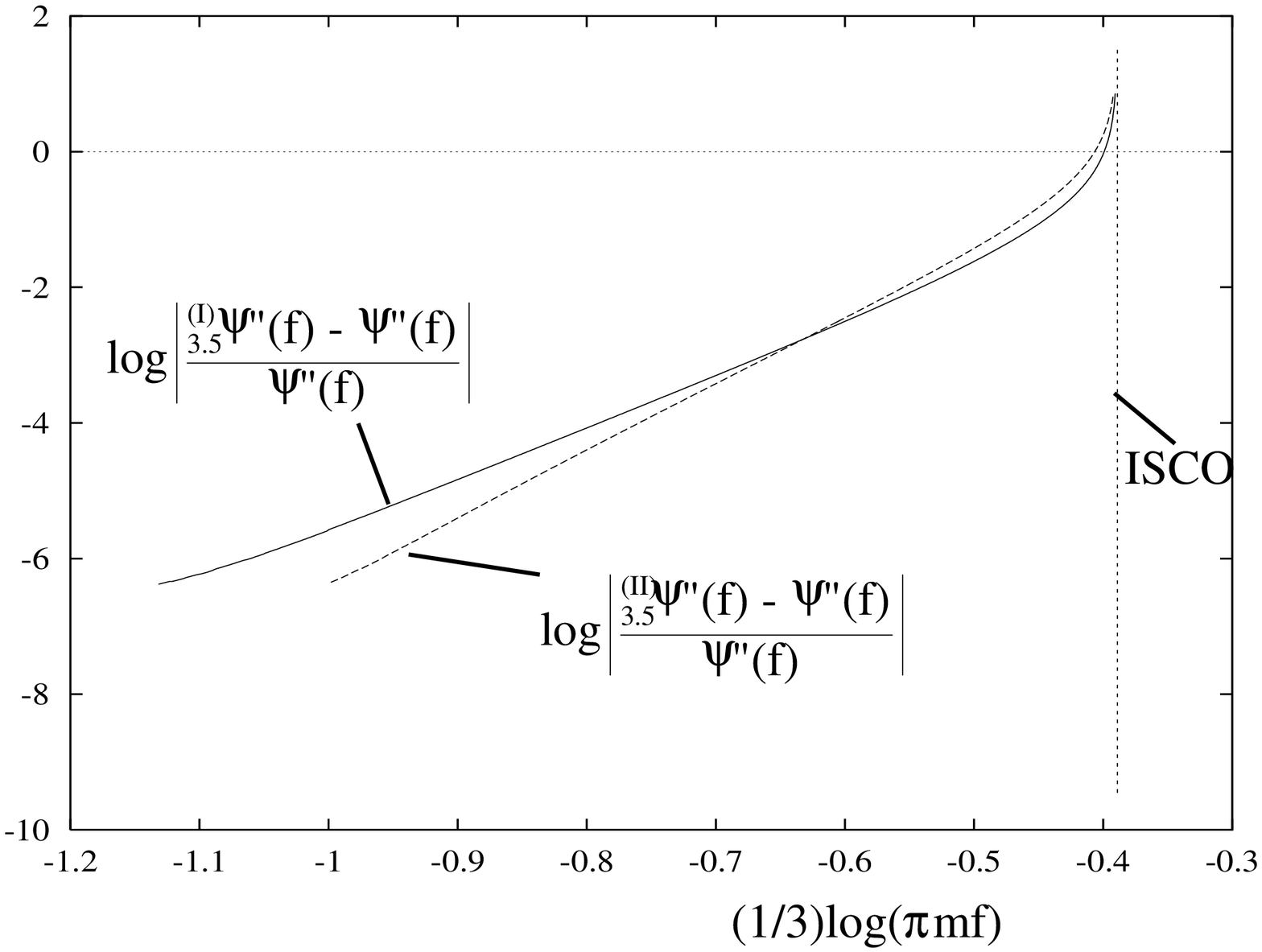}
\vspace{.4cm}
\caption{
The errors in the phase $\psi(f)$ when $E(f)$ and $F(f)$ are known
only up to post-3.5-Newtonian order; see caption of Fig.\ 1.
In this case method (II) is more accurate for most frequencies $f$.
}    
\label{fig-PN3.5}
\end{figure}

\begin{figure}
\epsfxsize=8.5cm
\epsfbox{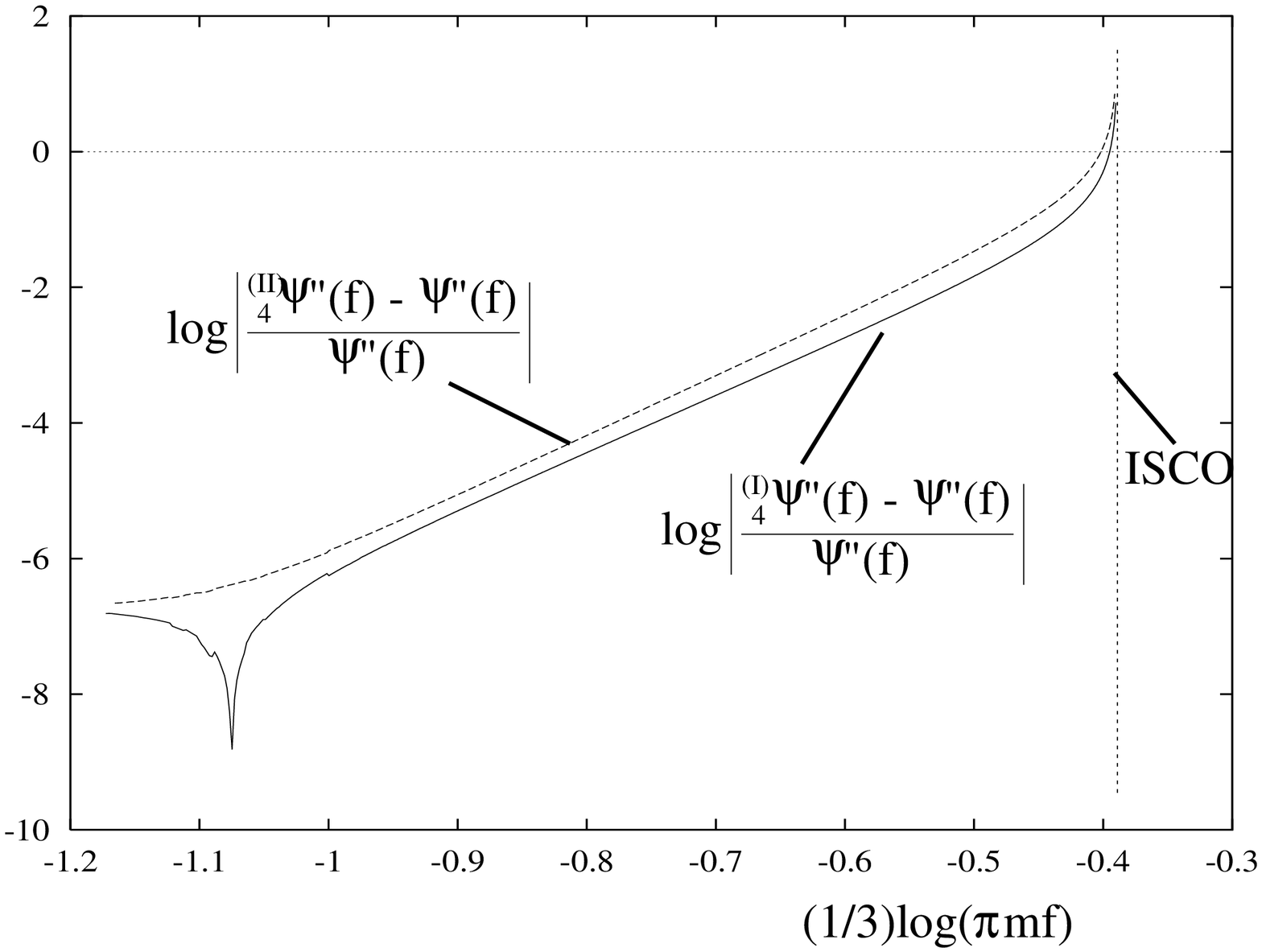}
\vspace{.4cm}
\caption{
The errors in the phase $\psi(f)$ when $E(f)$ and $F(f)$ are known
only up to post-4-Newtonian order; see caption of Fig.\ 1.
In this case method (I) is more accurate for all frequencies $f$.
}    
\label{fig-PN4}
\end{figure}

\begin{figure}
\epsfxsize=8.5cm
\epsfbox{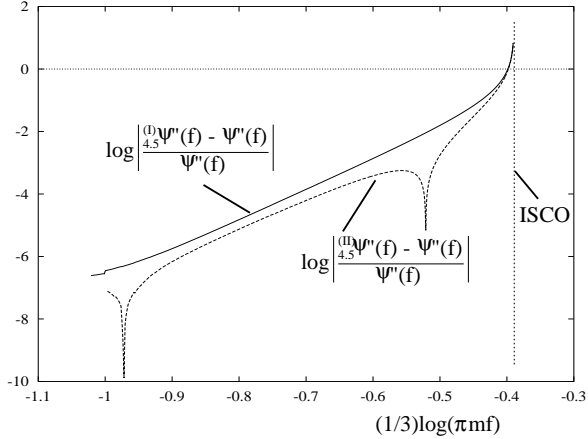}
\vspace{.4cm}
\caption{
The errors in the phase $\psi(f)$ when $E(f)$ and $F(f)$ are known
only up to post-4.5-Newtonian order; see caption of Fig.\ 1.
In this case method (II) is more accurate for all frequencies $f$.
}    
\label{fig-PN4.5}
\end{figure}

It is convenient to define the logarithmic relative errors
\begin{equation}
\ln \left| \frac{{}^{(I)}_{k/2}\psi'' (f) -\psi'' (f)}{\psi'' (f)} \right|
=\ln \left|
\frac{\left[\frac{E'(v)}{F(v)}\right]_k-\frac{E'(v)}{F(v)} }
     {\frac{E'(v)}{F(v)}} \right|
\end{equation}
and
\begin{equation}
\ln \left| \frac{{}^{(II)}_{k/2}\psi'' (f) -\psi'' (f)}{\psi'' (f)} \right|
=\ln \left|
\frac{\frac{\left[E'(v)\right]_k}{\left[F(v)\right]_k}-\frac{E'(v)}{F(v)} }
     {\frac{E'(v)}{F(v)}} \right| .
\end{equation}
These errors are shown in Figures \ref{fig-PN2.5} - \ref{fig-PN4.5}.
It can be seen that there is no systematic tendency for method (II)
to perform better than method (I). At post-2.5 and post-4-Newtonian
order method (I) does better than method (II) for all $v$, while at
post-3 and post-3.5-Newtonian order method (II) is more accurate than
method (I). We would expect the same trend to hold for general values
of the mass ratio $\eta$. 


\subsection{Overlaps of templates constructed by methods (I) and (II)
with the exact signal}

So far we have only considered how accurately the phase $\psi(f)$ is 
generated by methods (I) and (II). 
In this subsection we use the phases ${}^{(I)}_{k/2}\psi(f)$ and 
${}^{(II)}_{k/2}\psi(f)$ to construct gravitational wave templates and
then compute the templates' overlap with the exact waveform computed
from the exact $\psi(f)$. 

We use the restricted post-Newtonian approximation and neglect the
$m\ne2$ multipoles. Thus the Fourier 
transform ${\tilde h}(f)$ of the exact waveform
is given by \cite{accuracy_required}
\begin{equation}
\tilde{h}(f) \propto f^{-7/6} \mbox{e}^{i \psi(f) }  .
\end{equation}
Similarly we use methods (I) and (II) to construct the templates
\begin{equation}
{}^{(I)}_{k/2}\tilde{h}(f) 
\propto f^{-7/6} \mbox{e}^{i {}^{(I)}_{k/2}\psi(f) }  
\end{equation}
and
\begin{equation}
{}^{(II)}_{k/2}\tilde{h}(f) 
\propto f^{-7/6} \mbox{e}^{i {}^{(II)}_{k/2}\psi(f) }  .
\end{equation}

Next we compute Apostolatos' \cite{Apost-FF} 
fitting factor ($FF$) to determine the 
templates' accuracy. The fitting factor is the ratio of the
signal-to-noise ratio obtained with the imperfect template, to the
signal-to-noise ratio that a perfect template would yield.
The fitting factor can take values from zero to one, with unity indicating a
perfect template. It is obtained from the ambiguity function 
\begin{equation}
A=\frac{\left({}^{(I/II)}_{k/2}h,h\right)}
{\sqrt{ \left({}^{(I/II)}_{k/2}h,{}^{(I/II)}_{k/2}h\right) \left(h,h\right) }} 
\end{equation}
by maximizing over the template parameters, i.e.
\begin{equation}
FF=\mbox{max}_{\phi_c , t_c} A .
\end{equation}
Notice that we hold the masses fixed in the maximization procedure:
the templates and signal correspond to binaries of the same two
masses. Here we have introduced the inner product
\begin{equation}
(s,h) = 2\int_{0}^{\infty} 
\frac{\tilde{s}(f)^* \tilde{h}(f)+\tilde{s}(f)\tilde{h}(f)^* }{S_n (f)}df ,
\end{equation}
where $S_n (f)$ is the spectral density of the detector noise.
The noise curve $S_n (f)$ used here is the 
Cutler-Flanagan fit \cite{Cut-Flan-noise} 
for the advanced LIGO. The largest contribution to the overlaps 
comes from the frequency band between 40 Hz and 200 Hz. 

We compute the fitting factors for several different choices of $m_1$ 
and $m_2$ in order to get an indication of what might happen for
general mass ratios, even though our results apply strictly only to
the test mass limit $\eta \to 0$.
The resulting fitting factors are listed in tables 
\ref{1.4and1.4}, \ref{1.4and10} and \ref{10and10}. 
Examination shows
close agreement with the error plots of ${}^{(I)}_{k/2}\psi''(f)$ and
${}^{(II)}_{k/2}\psi''(f)$ in Figs. \ref{fig-PN2.5} -- \ref{fig-PN4.5}. 
At post-Newtonian orders
where ${}^{(I)}_{k/2}\psi''(f)$ has an error smaller than the error 
in ${}^{(II)}_{k/2}\psi''(f)$ everywhere (e.g. post-2.5), method (I)
always wins, and vice versa (e.g. post-4.5). On the other hand, at
post-Newtonian orders where the error lines cross (e.g. post-3.5), the
method with the smaller error in the $v$-region [$v=(\pi m f)^{1/3}$]
selected by the sensitive frequency band of the detector and the total
mass $m$ yields a larger fitting factor.

Again we find that there is no systematic tendency for method (II) to
be more accurate than method (I). Therefore method (II) does not lead
to a gain in accuracy when compared to method (I). Our conclusion
applies only to the limit $\eta \to 0$, but we do not anticipate a
different result for the general case.

\begin{table}
\caption{The fitting factor (FF) at post-$k/2$-Newtonian order 
for gravitational wave templates constructed by method (I) and (II).
A fitting factor of unity indicates a perfect template.
This table shows FF in the case of $m_1 = m_2 =1.4M_{\odot}$. 
\label{1.4and1.4}}
\begin{tabular}{|r|l|l|}
k & FF-method (I) & FF-method (II) \\
\hline
4  &  0.495960  &   0.710421 \\
5  &  0.963230  &   0.596752 \\
6  &  0.985540  &   0.981689 \\
7  &  0.995678  &   0.998342 \\
8  &  0.998731  &   0.997278 \\
9  &  0.999264  &   0.999851 \\
10 &  0.999661  &   0.999953 \\
11 &  0.999911  &   0.999964 \\
\end{tabular}
\end{table}

\begin{table}
\caption{The fitting factor (FF) for method (I) and (II) in the case of
$m_1 =1.4M_{\odot}$ and $m_2 =10M_{\odot}$.
\label{1.4and10}}
\begin{tabular}{|r|l|l|}
k & FF-method (I) & FF-method (II) \\
\hline
4  &  0.329050  &     0.871888 \\
5  &  0.787947  &     0.341547 \\
6  &  0.826911  &     0.872141 \\
7  &  0.926867  &     0.930399 \\
8  &  0.967903  &     0.919188 \\
9  &  0.978698  &     0.995905 \\
10 &  0.980767  &     0.990105 \\
11 &  0.983868  &     0.992731 \\
\end{tabular}
\end{table}

\begin{table}
\caption{The fitting factor (FF) for method (I) and (II) in the case of
$m_1 = m_2 =10M_{\odot}$.
\label{10and10}}
\begin{tabular}{|r|l|l|}
k & FF-method (I) & FF-method (II) \\
\hline
4  &  0.483132  &     0.908446 \\ 
5  &  0.895806  &     0.474364 \\ 
6  &  0.896781  &     0.976451 \\ 
7  &  0.959834  &     0.947991 \\ 
8  &  0.968440  &     0.948341 \\ 
9  &  0.972900  &     0.999764 \\ 
10 &  0.975185  &     0.994083 \\ 
11 &  0.987979  &     0.995887 \\ 
\end{tabular}
\end{table}

\acknowledgements

The research at Cornell was supported in part by NSF grant PHY 9722189
and by the Alfred P. Sloan foundation. The research at Guelph was
supported by the Natural Sciences and Engineering Research Council of
Canada.

\appendix

\section{Coefficients in expansions of energy and energy flux functions}
\label{sec-e_f_B}

\subsection{The coefficients $e_i$ and $f_i$ up to post-2.5-Newtonian order}

The coefficients in Eqs.\ (\ref{E_V}) and (\ref{F_V})
up to $2.5$ post-Newtonian order as given by 
Blanchet \cite{Blanchet} are:
\begin{equation}
e_{2}= -\frac{9+\eta}{12},
\end{equation}
\begin{equation}
e_{4}= -\frac{27-19\eta +\eta^2 /3}{8},        
\end{equation}
\begin{equation}
f_{2}=-\frac{1247}{336}-\frac{35\eta}{12},
\end{equation}
\begin{equation}
f_{3}=4\pi,
\end{equation}
\begin{equation}
f_{4}=-\frac{44711}{9072}+\frac{9271\eta}{504}+\frac{65\eta^2}{18},
\end{equation}
and
\begin{equation}
f_{5}=-\left(\frac{8191}{672}+\frac{535\eta}{24} \right) \pi.
\end{equation}

\subsection{The coefficients $e_i$, $f_i$ and $g_i$ up to
post-5.5-Newtonian order}

The remaining coefficients in Eqs.\ (\ref{E_V}) and (\ref{F_V}) 
have been given up to $5.5$ post-Newtonian order in the test mass 
limit in Ref.\ \cite{Tanaka-Tagoshi-Sasaki}.  These are
\begin{equation}
e_{6}=-\frac{675}{64},
\end{equation}
\begin{equation}
e_{8}=-\frac{3969}{128},
\end{equation}
\begin{equation}
e_{10}=-\frac{45927}{512},
\end{equation}
\begin{equation}
f_{6}= \frac{6643739519}{69854400}-\frac{1712\gamma}{105}
      +\frac{16\pi^2}{3}-\frac{3424 \log(2)}{105},
\end{equation}  
\begin{equation}
f_{7}= -\frac{16285\pi}{504},
\end{equation}
\begin{eqnarray}
f_{8}&=&-\frac{323105549467}{3178375200}+\frac{232597\gamma}{4410}
        -\frac{1369\pi^2}{126}   \nonumber \\
     & &+\frac{39931 \log(2)}{294}  -\frac{47385 \log(3)}{1568}, 
\end{eqnarray}
\begin{equation}
f_{9}= \frac{265978667519\pi}{745113600}-\frac{6848\gamma\pi}{105}
      -\frac{13696\pi \log(2)}{105},
\end{equation}
\begin{eqnarray}
f_{10}&=&-\frac{2500861660823683}{2831932303200}
         +\frac{916628467\gamma}{7858620}
         -\frac{424223\pi^2}{6804}   \nonumber \\
      & &-\frac{83217611 \log(2)}{11226600}  +\frac{47385 \log(3)}{196},
\end{eqnarray}
\begin{eqnarray}
f_{11}&=& \frac{8399309750401\pi}{101708006400}
         +\frac{177293\gamma\pi}{1176}    \nonumber \\
      & &+\frac{8521283\pi\log(2)}{17640}-\frac{142155\pi \log(3)}{784},
\end{eqnarray}
\begin{equation}
g_{6}=-\frac{1712}{105},
\end{equation}
\begin{equation}
g_{7}=0,
\end{equation}  
\begin{equation}
g_{8}=\frac{232597}{4410},
\end{equation}
\begin{equation}
g_{9}=-\frac{6848\pi}{105}, 
\end{equation}  
\begin{equation}
g_{10}=\frac{916628467}{7858620},
\end{equation}
and
\begin{equation}
g_{11}=\frac{177293\pi}{1176}.
\end{equation}  

\section{Analytic expressions for the coefficients in phase expansion
in various orders of approximation}
\label{sec-pqr}

\subsection{The coefficients $p_{j,n}$, $q_{j,n}$ and $r_{j,n}$ up to
          post-2.5-Newtonian order for general mass ratios}  

Here we list the analytic expressions for the coefficients $p_{j,n}$,
$q_{j,n}$ and $r_{j,n}$ up to post-2.5-Newtonian order for $\eta \neq
0$. 

\begin{eqnarray}
p_{2,2} = {{3715}\over {756}} + {{55\,\eta }\over 9}
\end{eqnarray}

\begin{eqnarray}
p_{3,2} = 0 
\end{eqnarray}

\begin{eqnarray}
p_{3,3} = -16\,\pi  
\end{eqnarray}

\begin{eqnarray}
p_{4,2} = p_{4,3} = {{5\,\left( 926521 + 1880368\,\eta  +
     905520\,{{\eta }^2} \right)} 
     \over {56448}} 
\end{eqnarray}

\begin{eqnarray}
p_{4,4} = {{15293365}\over {508032}} + {{27145\,\eta }\over {504}} + 
   {{3085\,{{\eta }^2}}\over {72}} 
\end{eqnarray}

\begin{eqnarray}
p_{5,2} = 0 
\end{eqnarray}

\begin{eqnarray}
p_{5,3} = p_{5,4} = {{20\,\left( 995 + 952\,\eta  \right) \,\pi }\over
{189}} 
\end{eqnarray}

\begin{eqnarray}
p_{5,5} = {{5\,\left( 7729 + 252\,\eta  \right) \,\pi }\over {756}}
\end{eqnarray}

\begin{eqnarray}
q_{2,n}=q_{3,n}=q_{4,n} = 0 
\end{eqnarray}

\begin{eqnarray}
q_{5,2} = 0 
\end{eqnarray}

\begin{eqnarray}
q_{5,3} = q_{5,4} = {{20\,\left( 995 + 952\,\eta  \right) \,\pi }\over
{63}} 
\end{eqnarray}

\begin{eqnarray}
q_{5,5} = {{38645\,\pi }\over {252}} + 5\,\eta \,\pi 
\end{eqnarray}

\begin{eqnarray}
r_{2,n}=r_{3,n}=r_{4,n}=r_{5,n}= 0 
\end{eqnarray}

\subsection{The coefficients $p_{j,n}$, $q_{j,n}$ and $r_{j,n}$ up to
post-5.5-Newtonian order in the test mass limit}
\label{sec-pqr_0}

Here we list analytic expressions for the remaining coefficients $p_{j,n}$,
$q_{j,n}$ and $r_{j,n}$ up to post-5.5-Newtonian order in the test
mass limit. 

\begin{eqnarray}
p_{6,2} = -{{5776858435}\over {9483264}} 
\end{eqnarray}

\begin{eqnarray}
p_{6,3} = -{{5776858435}\over {9483264}} - 320\,{{\pi }^2}
\end{eqnarray}

\begin{eqnarray}
p_{6,4}  = p_{6,5} = -{{37674179035}\over {85349376}} - 320\,{{\pi }^2}
\end{eqnarray}

\begin{eqnarray}
p_{6,6}&=&{{10817850546611}\over {4694215680}} - 
   {{6848\, \gamma }\over {21}}       \nonumber \\
&&-{{640\,{{\pi }^2}}\over 3} - 
   {{13696\,\log (2)}\over {21}} 
\end{eqnarray}

\begin{eqnarray}
p_{7,2} = 0 
\end{eqnarray}

\begin{eqnarray}
p_{7,3} = {{5680085\,\pi }\over {2352}} 
\end{eqnarray}

\begin{eqnarray}
p_{7,4} = {{152000375\,\pi }\over {63504}} 
\end{eqnarray}

\begin{eqnarray}
p_{7,5} = p_{7,6} = {{241249475\,\pi }\over {254016}} 
\end{eqnarray}

\begin{eqnarray}
p_{7,7} = {{77096675\,\pi }\over {254016}} 
\end{eqnarray}

\begin{eqnarray}
p_{8,2} = -{{7203742468445}\over {14338695168}} 
\end{eqnarray}

\begin{eqnarray}
p_{8,3} = -{{7203742468445}\over {14338695168}} - 
   {{43160\,{{\pi }^2}}\over {63}} 
\end{eqnarray}

\begin{eqnarray}
p_{8,4} = -{{499400855271485}\over {1161434308608}} - 
   {{43160\,{{\pi }^2}}\over {63}} 
\end{eqnarray}

\begin{eqnarray}
p_{8,5} = -{{499400855271485}\over {1161434308608}} - 
   {{47570\,{{\pi }^2}}\over {189}} 
\end{eqnarray}

\begin{eqnarray}
p_{8,6}&=& p_{8,7} = {{35381221594107617}\over {12775777394688}} - 
   {{1703440\, \gamma }\over {3969}}       \nonumber \\
&&-{{63110\,{{\pi }^2}}\over {567}} - {{3406880\,\log (2)}\over {3969}}
\end{eqnarray}

\begin{eqnarray}
p_{8,8}&=&{{2496799162103891233}\over {830425530654720}} - 
   {{36812\, \gamma }\over {189}}   
  -{{90490\,{{\pi }^2}}\over {567}}    \nonumber \\
&&-{{1011020\,\log (2)}\over {3969}} - 
   {{26325\,\log (3)}\over {196}} 
\end{eqnarray}

\begin{eqnarray}
p_{9,2} = 0 
\end{eqnarray}

\begin{eqnarray}
p_{9,3} = {{-6756514105\,\pi }\over {1185408}} - 640\,{{\pi }^3}
\end{eqnarray}

\begin{eqnarray}
p_{9,4} = {{-23087048755\,\pi }\over {3556224}} - 640\,{{\pi }^3}
\end{eqnarray}

\begin{eqnarray}
p_{9,5} = {{-971321608855\,\pi }\over {341397504}} - 640\,{{\pi }^3}
\end{eqnarray}

\begin{eqnarray}
p_{9,6}&=&{{121130241969551\,\pi }\over {18776862720}} - 
   {{27392\, \gamma \,\pi }\over {21}} - {{640\,{{\pi }^3}}\over 3} 
\nonumber \\
&&-{{54784\,\pi \,\log (2)}\over {21}} 
\end{eqnarray}

\begin{eqnarray}
p_{9,7}&=&p_{9,8}={{157063289889551\,\pi }\over {18776862720}} - 
   {{27392\, \gamma \,\pi }\over {21}}
\nonumber \\   
&&  - {{640\,{{\pi }^3}}\over 3} -{{54784\,\pi \,\log (2)}\over {21}} 
\end{eqnarray}

\begin{eqnarray}
p_{9,9}&=&{{90036665674763\,\pi }\over {18776862720}} - 
   {{13696\, \gamma \,\pi }\over {21}} - {{640\,{{\pi }^3}}\over 3} 
\nonumber \\   
&&-{{27392\,\pi \,\log (2)}\over {21}} 
\end{eqnarray}

\begin{eqnarray}
p_{10,2} = {{1796613371630183}\over {1070622572544}}
\end{eqnarray}

\begin{eqnarray}
p_{10,3} = {{1796613371630183}\over {1070622572544}} + 
   {{1240765\,{{\pi }^2}}\over {294}} 
\end{eqnarray}

\begin{eqnarray}
p_{10,4} = {{54094086068862461}\over {28906809458688}} + 
   {{11956093\,{{\pi }^2}}\over {2646}} 
\end{eqnarray}

\begin{eqnarray}
p_{10,5} = {{54094086068862461}\over {28906809458688}} + 
   {{458972531\,{{\pi }^2}}\over {338688}} 
\end{eqnarray}

\begin{eqnarray}
p_{10,6}&=&-{{4027802547645341665}\over {317974904045568}} + 
   {{650561605\, \gamma }\over {333396}}         \nonumber \\
&&+{{312163997\,{{\pi }^2}}\over {435456}} + 
   {{650561605\,\log (2)}\over {166698}} 
\end{eqnarray}

\begin{eqnarray}
p_{10,7}&=&-{{4027802547645341665}\over {317974904045568}} + 
   {{650561605\, \gamma }\over {333396}}          \nonumber \\
&&-{{138083683\,{{\pi }^2}}\over {435456}} + 
   {{650561605\,\log (2)}\over {166698}} 
\end{eqnarray}

\begin{eqnarray}
p_{10,8}&=&p_{10,9}=-{{23600127211067107843}\over {1878942614814720}} + 
   {{116990189\, \gamma }\over {166698}}          \nonumber \\
&&-{{181984501\,{{\pi }^2}}\over {3048192}} + 
   {{228376895\,\log (2)}\over {333396}}          \nonumber \\
&&+{{15716025\,\log (3)}\over {21952}}
\end{eqnarray}

\begin{eqnarray}
p_{10,10}&=&-{{1412206995432957982751}\over {126306697995878400}} + 
   {{6470582647\, \gamma }\over {27505170}}          \nonumber \\
&&+{{578223115\,{{\pi }^2}}\over {3048192}} + 
   {{53992839431\,\log (2)}\over {55010340}}          \nonumber \\
&&-{{5512455\,\log (3)}\over {21952}} 
\end{eqnarray}

\begin{eqnarray}
p_{11,2} = 0 
\end{eqnarray}

\begin{eqnarray}
p_{11,3} = {{-40905234824185\,\pi }\over {7169347584}} - 
   {{358720\,{{\pi }^3}}\over {189}} 
\end{eqnarray}

\begin{eqnarray}
p_{11,4} = {{-1456611391753955\,\pi }\over {193572384768}} - 
   {{358720\,{{\pi }^3}}\over {189}} 
\end{eqnarray}

\begin{eqnarray}
p_{11,5} = {{-1211268636338065\,\pi }\over {387144769536}} - 
   {{112990\,{{\pi }^3}}\over {189}} 
\end{eqnarray}

\begin{eqnarray}
p_{11,6}&=&{{41090763354419749\,\pi }\over {4258592464896}} - 
   {{7577740\, \gamma \,\pi }\over {3969}}     \nonumber \\
&&+{{15130\,{{\pi }^3}}\over {567}} - {{15155480\,\pi \,\log (2)}\over {3969}}
\end{eqnarray}

\begin{eqnarray}
p_{11,7}&=&{{50239568645429749\,\pi }\over {4258592464896}} - 
   {{7577740\, \gamma \,\pi }\over {3969}}      \nonumber \\
&&+{{15130\,{{\pi }^3}}\over {567}} - {{15155480\,\pi \,\log (2)}\over {3969}}
\end{eqnarray}

\begin{eqnarray}
p_{11,8}&=&{{3146788245124283189\,\pi }\over {276808510218240}} - 
   {{183628\, \gamma \,\pi }\over {189}}      \nonumber \\
&&-{{94390\,{{\pi }^3}}\over {567}} - 
   {{5572040\,\pi \,\log (2)}\over {3969}}      \nonumber \\
&&-{{26325\,\pi \,\log (3)}\over {49}} 
\end{eqnarray}

\begin{eqnarray}
p_{11,9}&=&p_{11,10}={{1846304168796859019\,\pi }\over {276808510218240}} - 
   {{449308\, \gamma \,\pi }\over {3969}}      \nonumber \\
&&-{{94390\,{{\pi }^3}}\over {567}} + 
   {{1241720\,\pi \,\log (2)}\over {3969}}      \nonumber \\
&&-{{26325\,\pi \,\log (3)}\over {49}} 
\end{eqnarray}

\begin{eqnarray}
p_{11,11}&=&{{1795505143426433771\,\pi }\over {276808510218240}} - 
   {{3558011\, \gamma \,\pi }\over {7938}}      \nonumber \\
&&-{{94390\,{{\pi }^3}}\over {567}} - 
   {{862549\,\pi \,\log (2)}\over {1134}}      \nonumber \\
&&-{{26325\,\pi \,\log (3)}\over {196}} 
\end{eqnarray}

\begin{eqnarray}
q_{6,2} = q_{6,3} = q_{6,4} = q_{6,5} =0 
\end{eqnarray}

\begin{eqnarray}
q_{6,6} = -{{6848}\over {21}} 
\end{eqnarray}

\begin{eqnarray}
q_{7,n} = 0 
\end{eqnarray}

\begin{eqnarray}
q_{8,2} = {{7203742468445}\over {4779565056}} 
\end{eqnarray}

\begin{eqnarray}
q_{8,3} = {{7203742468445}\over {4779565056}} + 
   {{43160\,{{\pi }^2}}\over {21}} 
\end{eqnarray}

\begin{eqnarray}
q_{8,4} = {{499400855271485}\over {387144769536}} + 
   {{43160\,{{\pi }^2}}\over {21}} 
\end{eqnarray}

\begin{eqnarray}
q_{8,5} = {{499400855271485}\over {387144769536}} + 
   {{47570\,{{\pi }^2}}\over {63}} 
\end{eqnarray}

\begin{eqnarray}
q_{8,6}&=&q_{8,7}=-{{37208950681636577}\over {4258592464896}} + 
   {{1703440\, \gamma }\over {1323}} \nonumber \\
&&+ {{63110\,{{\pi }^2}}\over {189}}
+ {{3406880\,\log (2)}\over {1323}}
\end{eqnarray}

\begin{eqnarray}
q_{8,8}&=&-{{2550713843998885153}\over {276808510218240}} + 
   {{36812\, \gamma }\over {63}} + {{90490\,{{\pi }^2}}\over {189}}      
\nonumber \\
&&+{{1011020\,\log (2)}\over {1323}} + {{78975\,\log (3)}\over {196}}
\end{eqnarray}

\begin{eqnarray}
q_{9,2} = q_{9,3} = q_{9,4} = q_{9,5} = 0 
\end{eqnarray}

\begin{eqnarray}
q_{9,6} = q_{9,7}=q_{9,8} = {{-27392\,\pi }\over {21}} 
\end{eqnarray}

\begin{eqnarray}
q_{9,9} = {{-13696\,\pi }\over {21}} 
\end{eqnarray}

\begin{eqnarray}
q_{10,2} = q_{10,3} = q_{10,4} = q_{10,5} = 0 
\end{eqnarray}

\begin{eqnarray}
q_{10,6} = q_{10,7} = {{650561605}\over {333396}} 
\end{eqnarray}

\begin{eqnarray}
q_{10,8} = q_{10,9} = {{116990189}\over {166698}} 
\end{eqnarray}

\begin{eqnarray}
q_{10,10} = {{6470582647}\over {27505170}} 
\end{eqnarray}

\begin{eqnarray}
q_{11,2} = q_{11,3} = q_{11,4} = q_{11,5} = 0 
\end{eqnarray}

\begin{eqnarray}
q_{11,6} = q_{11,7} = {{-7577740\,\pi }\over {3969}} 
\end{eqnarray}

\begin{eqnarray}
q_{11,8} = {{-183628\,\pi }\over {189}} 
\end{eqnarray}

\begin{eqnarray}
q_{11,9} = q_{11,10} = {{-449308\,\pi }\over {3969}} 
\end{eqnarray}

\begin{eqnarray}
q_{11,11} = {{-3558011\,\pi }\over {7938}} 
\end{eqnarray}

\begin{eqnarray}
r_{6,n}=r_{7,n} = 0 
\end{eqnarray}

\begin{eqnarray}
r_{8,2} = r_{8,3} = r_{8,4} = r_{8,5} = 0 
\end{eqnarray}

\begin{eqnarray}
r_{8,6} = r_{8,7} = {{1703440}\over {1323}} 
\end{eqnarray}

\begin{eqnarray}
r_{8,8} = {{36812}\over {63}} 
\end{eqnarray}

\begin{eqnarray}
r_{9,n}=r_{10,n}=r_{11,n} = 0 
\end{eqnarray}


\begin{references}

\bibitem{Wainstein}
L. A. Wainstein and V. D. Zubakov, {\it Extraction of signals from noise},
Prentice Hall, London, 1962.

\bibitem{Finn}
L. S. Finn, gr-qc/9903107.

\bibitem{Blanchet}
L. Blanchet, Phys. Rev. D {\bf 54}, 1417 (1996) (gr-qc/9603048).

\bibitem{accuracy_required}
E. Poisson, gr-qc/9801038;
S. Droz and E. Poisson, gr-qc/9712032; 
Phys. Rev. D {\bf 56}, 4449 (1997);

\bibitem{paperVI} 
E. Poisson, Phys. Rev. D {\bf 52}, 5719 (1995),
Addendum-ibid D {\bf 55}, 7980 (1997) (gr-qc/9505030).

\bibitem{Droz}
S. Droz, Phys. Rev. D {\bf 59}, 064030, (1999) (gr-qc/9806077). 

\bibitem{haris}
T. A. Apostolatos, Phys. Rev. D {\bf 54}, 2421 (1996).

\bibitem{note} 
This assumes that the binary's orbit is very nearly circular, which is
thought to be quite likely.  Nevertheless eccentric orbits are
possible in some scenarios for binary evolution \cite{PM}, and 
thus efforts are underway to construct families of templates with
non-zero eccentricity.  A related issue is that the current families
of templates need to be extended to incorporate modulations caused by 
spin-orbit coupling \cite{spin,haris} for the case of binaries
involving rapidly spinning companions.   

\bibitem{PM}
K. Martel and E. Poisson, gr-qc/9907006.

\bibitem{spin}
T.~A. Apostolatos, C.~Cutler, G.~J. Sussman, and K.~S. Thorne.
Phys. Rev. D, {\bf 49}, 6274 (1994).

\bibitem{DIS}
T.\ Damour, B.\ Iyer, and B.\ S.\ Sathyaprakash, Phys. Rev. D {\bf
57}, 885 (1998) (gr-qc/9708034).

\bibitem{Sathya}
B.S. Sathyaprakash, paper in preparation.

\bibitem{BSD}
R. Balasubramanian, B.S. Sathyaprakash, S.V. Dhurandhar, Phys. Rev. D
{\bf 53}, 3033 (1996); K. Kokkotas, A. Krolak, and G. Tsegas,
Class. Quant. Grav. {\bf 11}, 1901 (1994).

\bibitem{Kip}
K.S. Thorne, in {\it Black Holes and Relativistic Stars},
ed. R. M. Wald, University of Chicago press, Chicago, 1998
(gr-qc/9706079).  

\bibitem{DB} 
A. Buonanno and T. Damour, Phys. Rev. D {\bf 59}, 084006 (1999)
(gr-qc/9811091).

\bibitem{GRASP}
B. Allen, {\it GRASP: a data analysis package for gravitational wave
detection}, users manual for GRASP data analysis package release
1.9.4, p. 137.  GRASP is available at
http://www.lsc-group.phys.uwm.edu/$\tilde{\ }$ballen/grasp-distribution/ 


\bibitem{upperlimit}
B.\ Allen {\it et. al.}, Phys. Rev. Lett. {\bf 83}, 1498 (1999).

\bibitem{Droz-K-Poisson-O}
S. Droz, D. J. Knapp, E. Poisson, B. J. Owen,
Phys. Rev. D {\bf 59}, 124016 (1999)

\bibitem{Tanaka-Tagoshi-Sasaki}
T. Tanaka, H. Tagoshi and M. Sasaki,
Prog. Theor. Phys. {\bf 96}, 1087, (1996)

\bibitem{Apost-FF}
T. A. Apostolatos, Phys. Rev. D {\bf 52}, 605 (1996) 

\bibitem{Cut-Flan-noise}
C. Cutler, E. E. Flanagan, Phys. Rev. D {\bf 49}, 2658 (1994)



\end{references}
\end{document}